# Local magnetoresistance in Fe/MgO/Si lateral spin valve at room temperature

Tomoyuki Sasaki [1], Toshio Suzuki [2], Yuichiro Ando [3], Hayato Koike [1]
Tohru Oikawa [1], Yoshishige Suzuki [3], and Masashi Shiraishi [3]

1. Advanced Technology Development Center, TDK Corporation, Chiba, Japan
2. AIT, Akita Industrial Technology Center, Akita, Japan
3. Graduate School of Engineering Science, Osaka Univ. Toyonaka, Japan

Corresponding author: T. Sasaki (tomosasa@jp.tdk.com)

**Abstract**

Room temperature local magnetoresistance in two-terminal scheme is reported. By employing 1.6 nm-thick MgO tunnel barrier, spin injection efficiency is increased, resulting in large non-local magnetoresistance. The magnitude of the non-local magnetoresistance is estimated to be 0.0057 ohm at room temperature. As a result, a clear rectangle signal is observed in local magnetoresistance measurement even at room temperature. We also investigate the origin of local magnetoresistance by measuring the spin accumulation voltage of each contact separately.

Injection of spin polarized electron from a ferromagnetic electrode into a Si channel and transport of it thorough the Si channel are key issues for realization of Si based spin devices such as spin metal-oxide-semiconductor field-effect transistor (spin-MOSFET).[1] A lot of effort has been invested in order to establish these technologies in nearly a decade.[2-14] Non-local (NL) voltage measurement using a lateral spin valve device is one of the most potent way for demonstration of spin injection, accumulation and transport in a nonmagnetic channel.[15,16] Thus, spin related properties i.e., spin polarization, spin diffusion length have often been investigated by means of it. In this method, since the voltage measurement circuit is located outside of the current circuit as shown in the upper illustration of Fig. 1(a), voltage noise from the spin injector and the nonmagnetic channel is effectively excluded from the measured voltage, resulting in high sensitivity detection of electron spins. Recently, we have reported detection of NL magnetoresistance, NL-MR, at room temperature in Fe/MgO/Si lateral spin devices.[17] However, since most of practical spin devices proposed so far consist of source, drain, and gate,[1] detection of magnetoresistance by two-terminal measurement, i.e., local magnetoresistance, L-MR, is strongly desired.

In this study, we report detection of room temperature L-MR in Fe/MgO/Si lateral spin devices. Due to employing thick MgO tunnel barrier spin injection efficiency is increased up to twofold of that of previous study (see Ref.17), resulting in enhancement of spin accumulation in the Si channel. As a result, considerably large NL-MR is obtained. The magnitude of the NL magnetoresistance is estimated to be up to 0.0057 Ω at room temperature. As a result, a clear L-MR is observed even at room temperature.

As shown in Fig.1(a), the device consisted of a P doped (~$5\times10^{19}$ $cm^{-3}$) n-type Si channel with a width and thickness of 21 μm and 80 nm, respectively. The device was equipped with two ferromagnetic electrodes (contact 2 and 3) and two nonmagnetic electrodes (contact 1 and 4). The ferromagnetic and nonmagnetic electrodes consisted of an Fe/MgO bilayer and an Al layer, respectively. Thickness of the MgO tunneling barrier was 1.6 nm, which is thicker than that of previous studies.[17] Fig. 1(b) shows bias current, $I$, dependence of interfacial resistance area product of the contact 2 and 3. The resistance was decreased with increasing $|I|$, whose variation was pronounced compared with that of 0.8 nm-thick MgO barrier in our previous studies. The widths of the contact 2 and 3 were 0.35 and 2.0 μm, respectively, and the center-to-center distance between the contact 2 and 3, $d_L$, was 1.77 μm. The device fabrication method was described in detail elsewhere.[17] The conductivity of the Si channel at 8 and 300 K were measured to be $9.3\times10^4$ and $9.5\times10^4$ $\Omega^{-1}$ $m^{-1}$, respectively. Magnetoresistance measurements in NL and local scheme were employed under dc current injection. The measurements were repeated several times, and the data was averaged to realize more accurate analyses.

Figure 1(c) shows NL voltage as a function of magnetic field at $I$=5 mA measured at 8K. A clear rectangular hysteresis signal was observed. The magnitude of the signal was estimated to be 373 μV. As shown in Fig. 1(d), this hysteresis signal was observed even at room temperature, whose magnitude was estimated to be 28.5 μV. From these results NL-MR at 8 and 300 K were calculated to be 0.0746 and 0.0057 Ω, respectively, the largest value in our Si-based spin devices. Such marked enhancement in the NL-MR might be due to increment in thickness of the MgO tunnel barrier. In fact, spin polarization, $P$, calculated from NL-MR was increased and showed almost temperature independent behavior ($P$=0.075 at 5 mA). This behavior is obviously different from that of previous device with 0.8 nm-thick MgO layer, where the $P$ was drastically reduced from 0.04 to 0.01 when temperature is increased from 8 to 300 K [17]. L-MR at 8 and 300 K are shown in Fig. 1(e) and 1(f), respectively. The applied current was $I$=5 mA. As can be seen, a clear hysteresis behavior is observed even at 300 K. Local magnetoresistance, L-MR, at 8 and 300 K were calculated to be 0.66 and 0.075 Ω, respectively. This result is the first demonstration of magnetoresistance in local scheme in the semiconductor-based spin devices.

$I$ dependence of the L-MR was investigated, where a positive (negative) $I$ was defined as the case when spins were injected from the contact 2 (contact 3) into the Si channel. L-MR ratio as a function of $I$ is shown in the inset of Fig. 2(a). As can be seen, MR ratio is increased with increasing $|I|$. Here, we focus on the ratio between the L-MR and the NL-MR. L-MR/NL-MR as a function of bias current is summarized in the main panel of Fig. 2(a). In the spin diffusion model [18-20], the value of L-MR/NL-MR in our device geometry is expected to be 2, which is represented with a dashed line in Fig. 2(a). However, the experimental value of L-MR/NL-MR was considerably large. The deviation became pronounced with increasing $|I|$ and it reached up to 15 at $|I|$ =10 mA. In the spin diffusion model, L-MR in open geometry is expressed by using following equation;[18-20]

$$MR \equiv \frac{R_{AP} - R_P}{R_P} = \frac{\gamma^2}{1-\gamma^2} \frac{2Q_C}{2Q_C(1+Q_C)e^{\frac{d}{\lambda_N}} + \sinh\frac{d}{\lambda_N}} \quad .....(1),$$

where $\gamma$ is the spin polarization of injected current, $d$ is the distance between spin injector and detector, $\lambda_N$ is the spin diffusion length in the Si channel and $Q_c$ is the interface resistance divided by spin resistance of the Si channel. We neglected channel resistance, since it is so small compared to interface resistance. L-MR at $I$ = 10 mA as a function of $Q_c$ calculated by using eq.1 is shown in Fig. 2(b). The experimentally obtained value is also displayed in the same graph. The experimental value was obviously different from the theoretical one. This discrepancy was observed for all $I$ region from -10 to 10 mA. Here we emphasize that a clear L-MR was obtained in our devices, despite the substantial deviation from the optimum $Q_c$ condition. These facts indicate that the spin injection, accumulation and transport properties in our devices, which consist of a semiconductor channel and MgO tunnel barrier, are not entirely understood within the framework of the spin diffusion model.

To reveal the properties of local magnetoresistance in detail, further magnetoresistance measurements with different current-voltage scheme were carried out. As shown in Fig. 3(a), $I$ is applied between the contact 2 and 3, which is the same current configuration with that of Figs. 1(e) and (f), whereas voltage was measured between the contact 1 and 2, i.e., only voltage drop at the contact 2 was measured. Hereafter, this measurement scheme is referred to as “NL three-terminal” method. It should be noted that although detection of spin accumulation voltage with one ferromagnetic contact is well known as “three-terminal Hanle-effect measurement”, where magnetic fields along perpendicular to the film plane are applied, [6, 21] it demonstrates spin accumulation rather than spin transport. By contrast, the NL three-terminal method in this study can demonstrate spin transport because electron spins injected from the contact 3 are detected by the contact 2. As can be seen in Fig. 3(a), surprisingly, no clear signal was observed from the contact 2, unexpected behavior in the spin diffusion model. Conversely, when the voltage drop at the contact 3 was measured as shown in Fig. 3(b), a clear rectangle signal was observed. NL three-terminal resistance at the contact 3 is calculated to be 0.064 Ω. Figure. 3(c) shows $I$ dependence of the spin accumulation voltage measured by NL three-terminal method. The spin accumulation voltages measured by local and NL measurements are also displayed in the same graph. The spin accumulation voltage at contact 2 was obtained only at negative current ($I<0$), whereas that of the contact 3 was obtained only at positive current ($I>0$). This means that the spin accumulation voltage is obtained from the ferromagnetic contact only under the spin extraction condition. Here, we note that the spin accumulation voltage obtained by the NL three-terminal measurement under the spin extraction condition is almost the same magnitude with that of local measurements, indicating that the magnetoresistance in local scheme is caused by the spin accumulation voltage in one contact. The results shown in Figs. 2 and 3 suggest that in order to understand spin injection, accumulation, transport properties in lateral spin devices with Si channel and MgO tunnel barrier, the conventional spin diffusion model is not sufficient and an improved model should be established. A major difference between semiconductor-based spin device and metal-based spin device, whose spin transport properties can be understood in the framework of the spin diffusion model is the spin drift effect. In fact, modulation of spin transport length due to electric field was reported in our previous study [22, 23]. Using following spin drift-diffusion equation, [10, 24]

$$\frac{\partial S(x,t)}{\partial t} = D\frac{S^2(x,t)}{\partial x^2} - v\frac{\partial S(x,t)}{\partial x} - \frac{S(x,t)}{\tau_s}, \quad .... (2)$$

where $S(x,t)$ is spin density at a position ($x$) and time ($t$), $D$ is the diffusion constant, $\tau_s$ is the spin relaxation time and $v$ is the spin drift velocity, spin accumulation voltage, $V$, is expressed as;

$$\mathrm{V} = V_0 exp\left(\frac{vd_L}{2D}\right)\frac{1}{\sqrt{v^2+\frac{4D}{\tau_s}}}\exp(-\frac{d_L}{2D}\sqrt{v^2+\frac{4D}{\tau_s}}), \quad .... (3)$$

where $V_0$ is constant value independent of the electric field. Using eq. (3), L-MR/NL-MR value is roughly estimated to be 7 at $I$ = 10 mA ( electric field in Si channel is estimated to be 0.8 MV/m) and the ratio between spin drift length and spin diffusion length is estimated to be 2.1.[22, 23] Therefore, contribution of spin drift effect should be taken into account.[25] However, there is still non-negligible discrepancy between experimental value and theoretical one in Fig. 2(a). Moreover, the results in Fig. 3 cannot be explained even if the spin drift is took into account. Rigorous calculation approach which takes into account other factors e.g., effect of a tunnel barrier on spin detection, bias dependence of spin polarization, and so on might be needed.

In summary, we have reported two-terminal magnetoresistance in Si-based lateral spin valve at room temperature. Due to employing thick MgO tunnel barrier spin injection efficiency was drastically increased resulting in enhancement of magnetoresistance. However, the magnitude of magnetoresistance was larger than that theoretically expected using conventional spin diffusion model. NL three-terminal measurements revealed that spin accumulation voltage was detected only under the spin extraction condition. This means that local magnetoresistance is caused by spin accumulation voltage at one ferromagnetic contact under spin extraction condition. In order to understand spin injection, accumulation, transport properties in lateral spin devices with Si channel and MgO tunnel barrier, the conventional spin diffusion model is not sufficient and an improved model should be established.

**Figure Captions**

**Figure 1 (Color online)**

(a) A schematic of the Si-based lateral spin valve with four terminals. An external magnetic field is applied along ±*y* direction. Contacts 2 and 3 consist of Fe/MgO bilayer and contacts 1 and 4 consist of Al layer. Current-voltage scheme for nonlocal and local magnetoresistance measurement are shown in upper and lower figure, respectively. (b) Resistance area product of contact 2 and 3 as a function of bias current at 300 K. The RA value measured at $I = 100$ μA is estimated to be 100 kΩ μm$^2$, more than twenty fold of that of 0.8 nm-thick MgO barrier reported in ref. 17. Non-local magnetoresistance measured at (c) 8 K and (d) 300 K, where the bias current was set to +5 mA. Local magnetoresistance measured at (c) 8 K and (d) 300 K. The bias current was set to +5 mA.

**Figure 2**

(a) Ratio of local magnetoresistance, L-MR, to nonlocal magnetoresistance, NL-MR, as a function of bias current. The solid circles are experimental data and dot-line is a theoretically expected value in our device geometry. (b) MR ratio calculated by using the spin diffusion model as a function of $Q_c$, where $Q_c$ is the interface resistance divided by spin resistance of the Si channel. The bias current was set to 10 mA. An open circle is the experimental data measured by the local scheme.

**Figure 3**

Spin accumulation voltages of (a) contact 2 and (b) contact 3 measured by the non-local 3-terminal measurements under +5 mA at 300 K. The magnetic field was applied parallel to the *y*-direction. The insets show the schematic illustrations of current voltage scheme. (c) Spin accumulation voltage measured by local, nonlocal and nonlocal three-terminal measurements as a function of bias current at 300 K. The bias current was set to 5 mA.

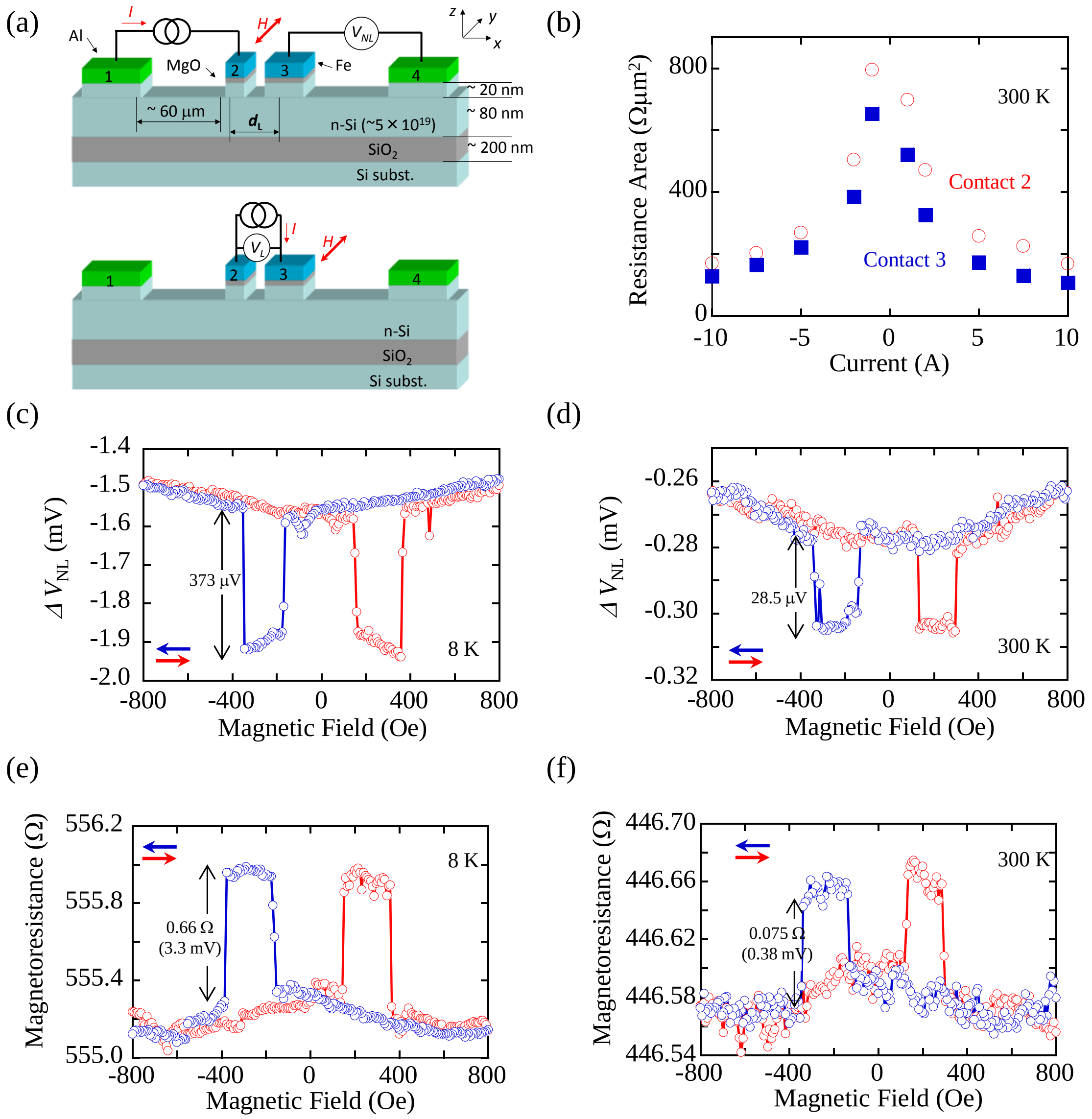

Fig. 1 T. Sasaki et al.

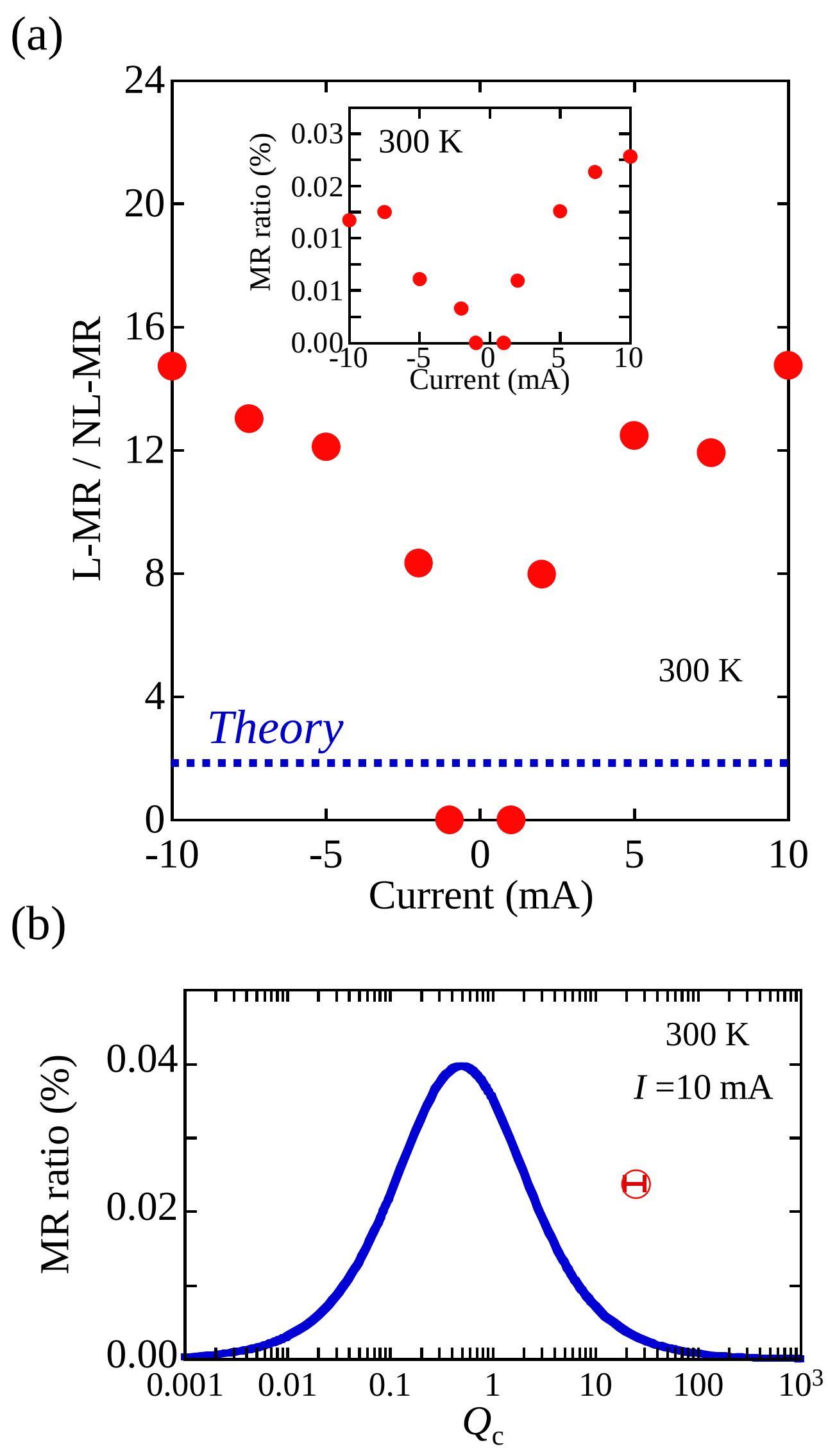

Fig. 2 T. Sasaki et al.

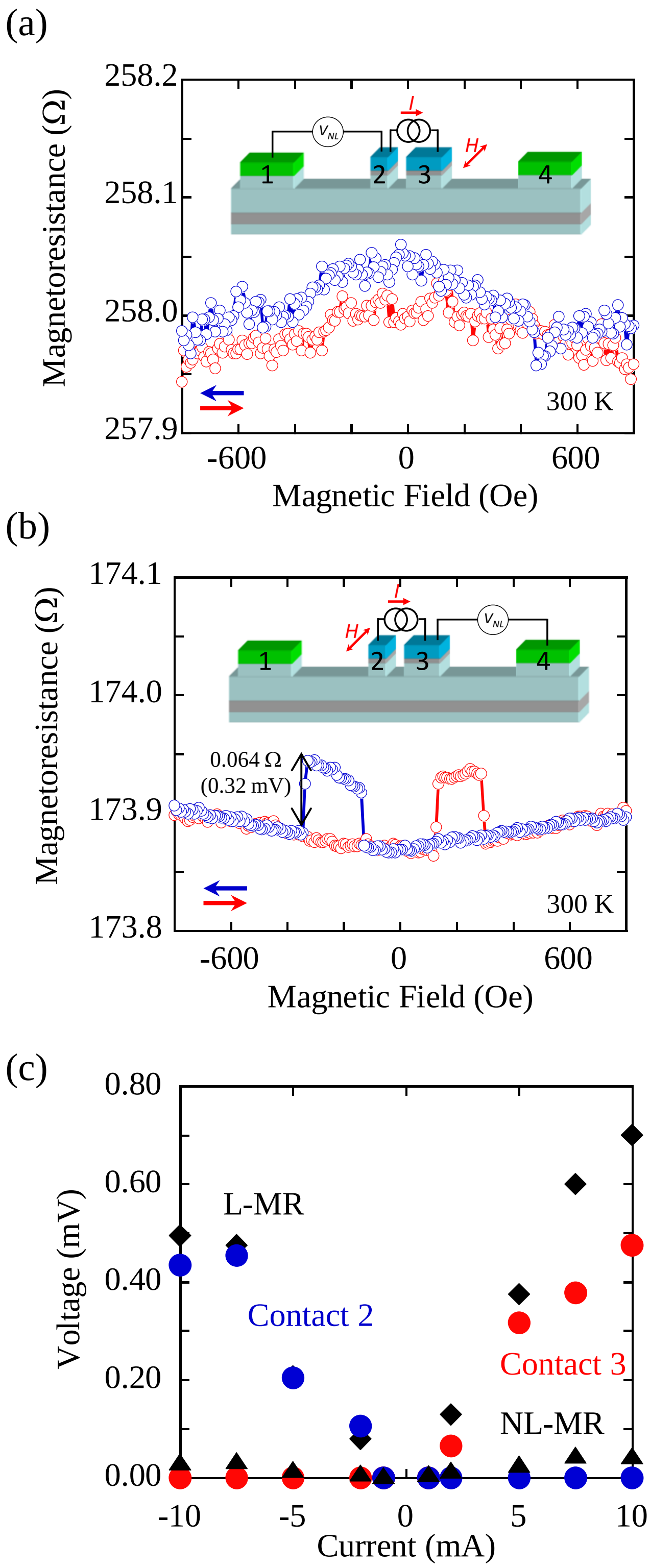

Fig. 3 T. Sasaki et al.